\documentstyle[11pt,newpasp,twoside,epsf]{article}
\def\edcomment#1{\iffalse\marginpar{\raggedright\sl#1\/}\else\relax\fi}
\reversemarginpar

\begin{document}
\title{The different clustering of red and blue galaxies: a robust signal
from $\omega(\theta)$}
 \author{R\'emi A. Cabanac}
\affil{E.S.O., Alonso de Cordoba 3107, casilla 19001, Vitacura, Santiago 19, 
Chile}
\author{Val\'erie de Lapparent}
\affil{I.A.P., 98 bis Bld Arago, 75014 Paris, France}
\author{Paul Hickson}
\affil{U.B.C., Dept. of physics and astronomy, Hennings Bld, Vancouver, V6T 1Z4, Canada}
\begin{abstract}
A sample of $\sim$ 20,000 galaxies covering 0.76 deg$^2$ were observed with the 
CFHT-UH8K up to $V<23.5$ and $I<22.5$. The angular correlation analysis of the 
red selected sample ($V-I>1.4$) shows a stronger amplitude than the blue 
selected sample at all cutoff magnitudes. This effect could be explained either
by luminosity selection effects or by a true color segregation in which red objects are 
preferably found in denser regions. If the latter is true, and assuming that 
red galaxies are dominated by an older population of stars, this measurement 
supports the idea that galaxies evolve faster in the harsh environment of 
dense clusters than in the field, at fiducial redshifts of z$\sim$0.5.
\end{abstract}
\noindent{\bf{Observations:}} We observed 4 contiguous CFHT-UH8K fields, 
covering $0.5^\circ \times 1.5^\circ $ near the north galactic pole.
The observations were complete to $I=22.75$ and $V=23.75$ but, in order to avoid
systematic biases, we measure the correlation functions to $I<22.5$ and $V<23.5$. 
Because of the color selection, faint objects redder than $V-I>1$ are missed 
near the limit of the I catalogue. \\
{\bf{Analysis and Results:}} The imaging reduction, photometry (uncertain to
$\pm0.05$ magnitude in V) and astrometry (uncertain to 0.5\arcsec~over the 
entire field) is fully described in Cabanac et al. (2000).\\
The angular correlation function $\omega(\theta)=A_\omega \theta^{-\delta}$
measured over the entire sample of galaxies does not reveal any variation of 
the slope $\delta=0.8$ with increasing limiting magnitude. 
The angular correlation function (with a fixed slope $\delta=-0.8$) is 
measured separately on two samples of galaxies discriminated by their $V-I$ 
color (7259 blue galaxies with $V-I<1.4$; 8986 red galaxies with $V-I>1.4$),
in the intervals $17<I<21, 17<I<21.5, 17<I<22.0, 17<I<22.5, I>17$ and
to the limits $I<22.5, V<22.5$ (Figure 1).\\
Assuming the luminosity function of the CFRS, we transpose the amplitude of 
the angular correlation $A_\omega$ into the amplitude of the spatial 2-point 
correlation function $r_0 = A_\omega^{1/1.8} / (3.68 B[D(z),N(z),z])^{1/1.8}$ 
with $B[D(z),N(z),z]$ defined in Cabanac et al. (2000).\\
Fiducial redshifts for a $\Lambda = 1/3$ flat universe are $z = 0.38, 0.44, 
0.51, 0.56$.\\
Red galaxies always show a higher amplitudes than blue objects for all 
magnitude intervals (Figure 1).\\
\begin{figure}
\plotfiddle{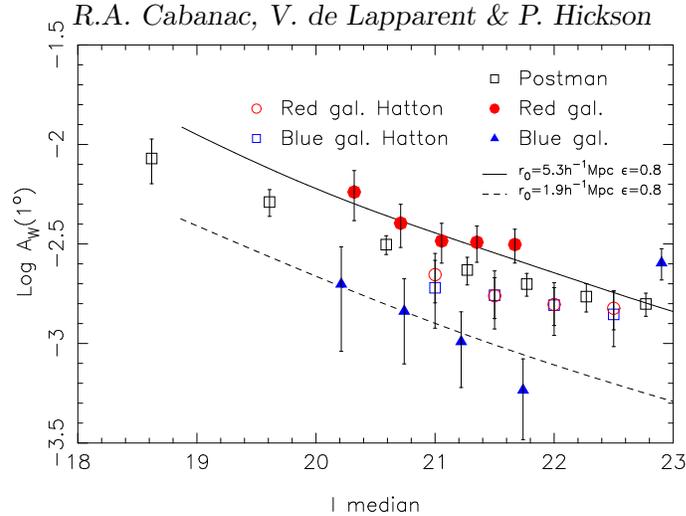}{5cm}{0}{110}{110}{-110}{-25}
\caption{The amplitude of correlation $A_\omega$ is shown for red galaxies 
(circles) and blue galaxies (triangles). The lines are best-fit 
amplitudes $r_0$ of the spatial 2-point correlation function (cf text). 
Postman et al. (1998) measurements are shown as black open squares. The grey 
open symbols (from a CDM N-body simulation [courtesy of S. Hatton]) show no 
difference between red and blue samples. The error bars are proportional to 
the number of galaxies in each correlation measurement. The last triangle 
to the right gives an idea of the systematic error caused by incomplete 
samples ($I > 17$), as un-physical correlations are introduced by zero-point 
errors, different sensitivities of the frames, etc.}
\end{figure}
{\bf{Discussion:}} This result is consistent with local measurements where 
bulge-dominated galaxies have a higher correlation length than disk-dominated 
galaxies (e.g. Loveday et al., 1995). It strongly confirms previous 
observations of Neuschafer et al. (1995), Lidman \& Peterson (1996) and Roche 
et al. (1996).\\
This result can be explained if blue galaxies are intrinsically fainter than 
red galaxies and are present in the field as well as in the clusters, whereas 
red galaxies are larger, brighter and tend to concentrate in the clusters.
A possible philogenic explanation could be that galaxies tend to evolve more 
rapidly in clusters than in the field. Physical processes such as ram pressure,
tidal effects, galaxy harassment, near misses would trigger an earlier star 
formation in clusters and deprive cluster galaxies of their gas whereas 
galaxies of the field would sustain a regular but weaker star formation,
still on-going nowadays.

\end{document}